\documentclass{emulateapj}

\shorttitle{Binary quasar FIRST\,J1643+3156}
\shortauthors{Kunert-Bajraszewska and Janiuk}

\begin{document}


\title{Discovery of the disturbed radio morphology in the interacting binary quasar
FIRST\,J164311.3+315618} 

\author{Magdalena Kunert-Bajraszewska$^{1}$}
\affil{$^{1}$Toru{\'n} Centre for Astronomy, N. Copernicus University,
87-100 Toru{\'n}, Poland}
\email{magda@astro.uni.torun.pl}

\and
\author{Agnieszka Janiuk$^{2}$}
\affil{$^{2}$
Center for Theoretical Physics, Polish Academy of Sciences, Al. Lotnikow
32/46, 02-668 Warsaw, Poland}

\begin{abstract}
We report the high resolution radio observations and their analysis of a
radio-loud compact steep spectrum (CSS) quasar FIRST\,J164311.3+315618, one of the members of 
a binary system. The second component of the system is a radio-quiet
AGN.  
The projected separation of this pair is 2.3$\arcsec$ (15\,kpc) and it is
one of the known smallest separation binary quasars. The multi-band images 
of this binary system made 
with the Hubble Space Telescope showed that the host galaxy of the radio-loud quasar is highly
disturbed.
The radio observations presented here were made with the multi-element radio linked
interferometer network (MERLIN) at 1.66\,GHz and 5\,GHz. 
We show that the radio morphology of FIRST\,J164311.3+315618 is complex
on both frequencies and exhibits four components, which 
indicate on the intermittent activity with a possible rapid
change of the jet direction and/or restart of the jet due to the interaction with the companion.
The radio components that are no longer powered by the jet can
quickly fade away. We suggest that this makes the potential distortions of the radio structure 
to be short-lived phenomena. On the other hand, our numerical simulations show that the
influence of the companion can lead to the prolonged current and future 
activity. 
FIRST\,J164311.3+315618 is an unusual and statistically very rare low redshift binary
quasar in which probably the first close encounter is just taking place. 
\end{abstract}


\section{Introduction}
It has been established that most galaxies host a super-massive
black hole \citep{magorrian}, and the next open issue is   
the mechanism that triggers gas accretion and nuclear activity.
It is likely that active galactic nuclei can pass
through subsequent stages of higher and lower activity. Intermittent 
behavior of this type can be caused by minor mergers or
instabilities in the accretion flow. 
On long timescales, up to $10^{8}$ years, the merger
activity is broadly considered as the cause of an enhanced accretion
flow and the main provider of material for the black hole growth
\citep{barnes, volonteri, dimateo}. 

Close binary QSOs presumably represent a very early stage of a merger of  
galaxies. The merger will
likely produce a supermassive binary black hole \citep{bege}.  
Based on optical surveys, only about 0.1\% of all quasars observed
have one, or more, nearby quasar companion at the same
redshift \citep[e.g.][]{kochanek99, hewett, foreman}. 
These may either be gravitational lenses, true quasar pairs, or chance
alignments. Arguments weighted after of binary quasar classification are:
lack of a detectable lens, different quasar spectra and/or properties like
the case when one object is radio-quiet and the other radio-loud.  
There is a growing number of identifications of physical pairs of
quasars or as good candidates \citep{kochanek99, hennawi, liu}.
Among them the highest-redshift binary known LBQS\,0015+0239 at {\it z}=2.45
\citep{impey}, LBQS\,1429-0053 \citep{faure} and UM\,425 \citep{mathur,
aldcroft} with similar optical spectra but lack of a candidate lensing
system, and
Q\,2345+007 \citep{green2002} with different optical and X-ray spectra but
no lens candidate found in optical and X-ray band.  
The smallest-separation
known binaries are LBQS\, 0103-2753 \citep{jun01} with separation
0.3$\arcsec$ and very recently discovered SDSS J1536+0441
\citep{boroson, decarli} with separation 1$\arcsec$.  
While most of the binaries are $O^{2}$ pairs (both quasars are radio faint),
there are four known $O^{2}R$ systems (one quasar is radio-loud, and the
other is radio-quiet): PKS\,1145-071 \citep{djorgovski}, MGC\,2214+3550
\citep{munoz}, Q1343+2640 \citep{crampton}, and FIRST\,J164311.3+315618
\citep{broth} studied in the present paper. The existence of two images with
extremely different flux ratios in the optical and the radio strongly
favors the binary quasar classification. 

Most of the binaries are high redshift objects with separation
3$\arcsec$-10$\arcsec$, which corresponds to distances of $\sim 10-80$\,kpc
\citep{mortlock}. According to \citet{jun01} the observed
3$\arcsec$-10$\arcsec$ binary QSOs   
mostly represent galaxy pairs undergoing the loop following the first close
encounter. The still open question is if the host galaxies of the binary quasars
interact in the process of merging. If so, 
we should observe morphological and kinematical distortions in these systems
and their character should depend on the evolutionary stage of the pair. In
the late phase of their encounter we might expect compact, morphologically
highly disturbed system with a pair of active supermassive black holes at
its center. A few binary supermassive black holes have been observed so
far: NGC\,6240
\citep{komossa}, Arp\,299 \citep{ballo}, 0402+379
\citep{rodri} and COSMOS J100043.15+020637.2 \citep{comerford}, and cluster
members 3C75 \citep{owen} and J0321-455 \citep{klamer}.
Strong evidence for the interaction of the host
galaxies of the binary AGN system SDSS J1254+0846 (separation 3.8$\arcsec$) has
been provided through the observed distortion of the optical light from
one of the host galaxies, showing obvious tidal tails \citep{green2010}.
Here
we report that radio-loud/radio-quiet binary quasar system associated
with the radio source FIRST\,J164311.3+315618 (hereafter FIRST\,J1643+3156) 
shows disturbed radio morphology possibly indicating that
the two quasars are in the process of merging.
The distortions of the host galaxy of the radio-loud quasar of this 
system have been also discovered \citep{martel}.

The radio-loud/radio-quiet binary quasar associated with the radio source
FIRST\,J1643+3156
has been classified by \citet{broth} based on the optical observations. It
is a small separation, 2.3$\arcsec$ (15\,kpc), quasar
pair with redshift $z=0.586$ and greatly discrepant optical and radio flux ratios ($O^{2}R$
pairs).
On this basis the gravitational lens hypothesis has been ruled out
\citep{broth}.
The radio-loud component of this pair is characterized by strong, narrow
emission line spectrum and X-ray emission. In the radio-quiet one the
starburst has been triggered \citep{broth}. There is a very small difference in the
quasars' redshifts: $z=0.5867$ for radio-loud component, and $z=0.5862$ for
radio-quiet component. A crosscorrelation
analysis indicates that radio-loud source is redshifted relative to
the radio-quiet one by the order of $V= 300$ km s$^{-1}$ \citep{broth}. 
The radio emission of the radio-quiet component is undetected. 
The radio-loud component of the binary quasar, FIRST\,J1643+3156, is a Compact Steep
Spectrum
(CSS) radio source with spectral index
$\alpha_{1.4\mathrm{GHz}}^{4.85\mathrm{GHz}}=0.82$. It has a total flux
density of $S_{1.4\,{\rm GHz}}=113~{\rm mJy}$, what gives the moderate luminosity
of
log$L_{1.4\mathrm{GHz}}$=26.20~W~${\rm Hz^{-1}}$. The black hole mass of the
radio-loud
component has been estimated \citep{shen10} to be $M_{\rm BH} = 3\times
10^{8} M_{\odot}$, the bolometric luminosity assumes
log$L_{bol}$=45.736~${\rm ergs\,s^{-1}}$, and the Eddington ratio is
$L_{bol}/L_{Edd} = 0.16$
(Table~\ref{table1}).

In this paper we present high resolution radio observations of the
radio-loud component of the binary system and their analysis. 
The observations were made with the multi-element radio linked
interferometer network (MERLIN) at 1.66\,GHz and 5\,GHz.
The unusual
complex morphology of FIRST\,J1643+3156 is discussed as the consequence of the
quasars interaction and/or the restart of the activity of the radio-loud one.

\begin{table}
\begin{center}
\caption[]{Basic parameters of FIRST\,J1643+3156}
\begin{tabular}{@{}l r c@{}}
\hline
\hline
Parameter & Value & Ref\\
\hline
Other source name (B1950)     & 1641+320 &\\
RA (J2000) extracted from FIRST & $16^{\rm h}43^{\rm m}11.35^{\rm s}$ &\\
Dec (J2000) extracted from FIRST & $+31^{\rm o}56$\arcmin$18.00$\arcsec$$ &\\
Redshift {\it z}& 0.586&(2)\\
Total flux density $S_{1.4\,{\rm GHz}}$~(mJy)&$112\pm 6$&(1) \\
log$L_{1.4\mathrm{GHz}}$~(W~${\rm Hz^{-1}}$)     & 26.20&\\
Total flux density $S_{4.85\,{\rm GHz}}$~(mJy)& $41\pm 7$&(3) \\
log$L_{4.85\mathrm{GHz}}$~(W~${\rm Hz^{-1}}$)     & 25.76&\\
Spectral index $\alpha_{1.4\mathrm{GHz}}^{4.85\mathrm{GHz}}$&$0.82$&\\
Largest Linear Size ($h^{-1}~{\rm kpc}$) & 10.56&(5)\\
0.1-2.4\,keV flux (${\rm ergs\,s^{-1}\,cm^{-2}}$)& $1.1\times10^{-12}$&(2)\\
log$L_{bol}$ (${\rm ergs\,s^{-1}}$)& $45.736\pm 0.008$&(4)\\   
log$M_{BH}/M_{\odot}$ & $8.43\pm 0.07$&(4)\\
Eddington ratio & 0.16&(4)\\
Projected separation & & \\ 
of the optical components & 2.3$\arcsec$&(2)\\
\hline
\end{tabular}
\end{center} 
Note. Spectral index defined as ($S\propto\nu^{-\alpha_{r}}$). References:
(1) \citet{becker95}, (2) \citet{broth}, (3) \citet{gregory}, (4)
\citet{shen10}, (5) this paper.
\label{table1}
\end{table}

\section{Radio observations}
\label{obs}

The binary quasar FIRST\,J1643+3156 belongs to the
sample of 44 Low Luminosity Compact (LLC) objects we observed and analyzed
in \citet{kun10a}. 
The snapshot 1.66\,GHz and 5\,GHz MERLIN observations of the
FIRST\,J1643+3156 were undertaken 
in 2007 and 2009.
The target source together with its associated phase reference
source was observed for $\sim$60\,min including telescope drive times.
The initial data reduction of raw MERLIN data   
was made using local d-programs and AIPS-based PIPELINE procedure developed
at Jodrell Bank Observatory ($http://www.merlin.ac.uk/user_{-}guide/$). 
OQ208 was used as the point source or baseline
calibrator and 3C\,286 as the flux and polarization calibrator.  
Further cycles of phase self-calibration and imaging using the NRAO AIPS
software was then used to produce the final total intensity (I) and polarization 
intensity (P) images. The polarization was detected only in component
W2.
The flux densities of the main components of
the target source were then measured, by fitting Gaussian models, using AIPS task JMFIT
(Table~\ref{component}). 
The position of the optical counterpart of target source is marked 
with a cross in the maps and was taken from the Sloan Digital Sky Survey
(SDSS/DR7).

Throughout the paper, we assume a cosmology with
${\rm
H_0}$=71${\rm\,km\,s^{-1}\,Mpc^{-1}}$, $\Omega_{M}$=0.27,
$\Omega_{\Lambda}$=0.73.

\begin{table}
\begin{center}
\caption[]{Flux densities of FIRST\,J1643+3156 principal components from the
1.66\,GHz and 5\,GHz MERLIN images}
\begin{tabular}{ccccc}
\hline
Compo- & ${\rm S_{1.66\mathrm{GHz}}}$&$\theta_{min}\times \theta_{maj}$
&${\rm
S_{5\mathrm{GHz}}}$&$\theta_{min}\times \theta_{maj}$ \\
 nents          &   mJy &  arcs           & mJy& arcs\\ 
\hline    
C & $14\pm 0.6$ &$0.13\times 0.23$  & $7\pm 0.3$&$0.01\times 0.02$  \\
E & $18\pm 0.8$ &$0.22\times 0.31$  & $-$& $-$\\
W1 &$22\pm 0.6$ &$0.13\times 0.23$  & $2\pm 0.3$&$0.02\times 0.03$  \\
W2 &$32\pm 0.7$ &$0.17\times 0.31$  & $3\pm 0.3$&$0.02\times 0.07$  \\
\hline
\end{tabular}
\end{center} 
\label{component}
\end{table}

\begin{figure*}
\centering
\includegraphics[width=17cm,height=11cm]{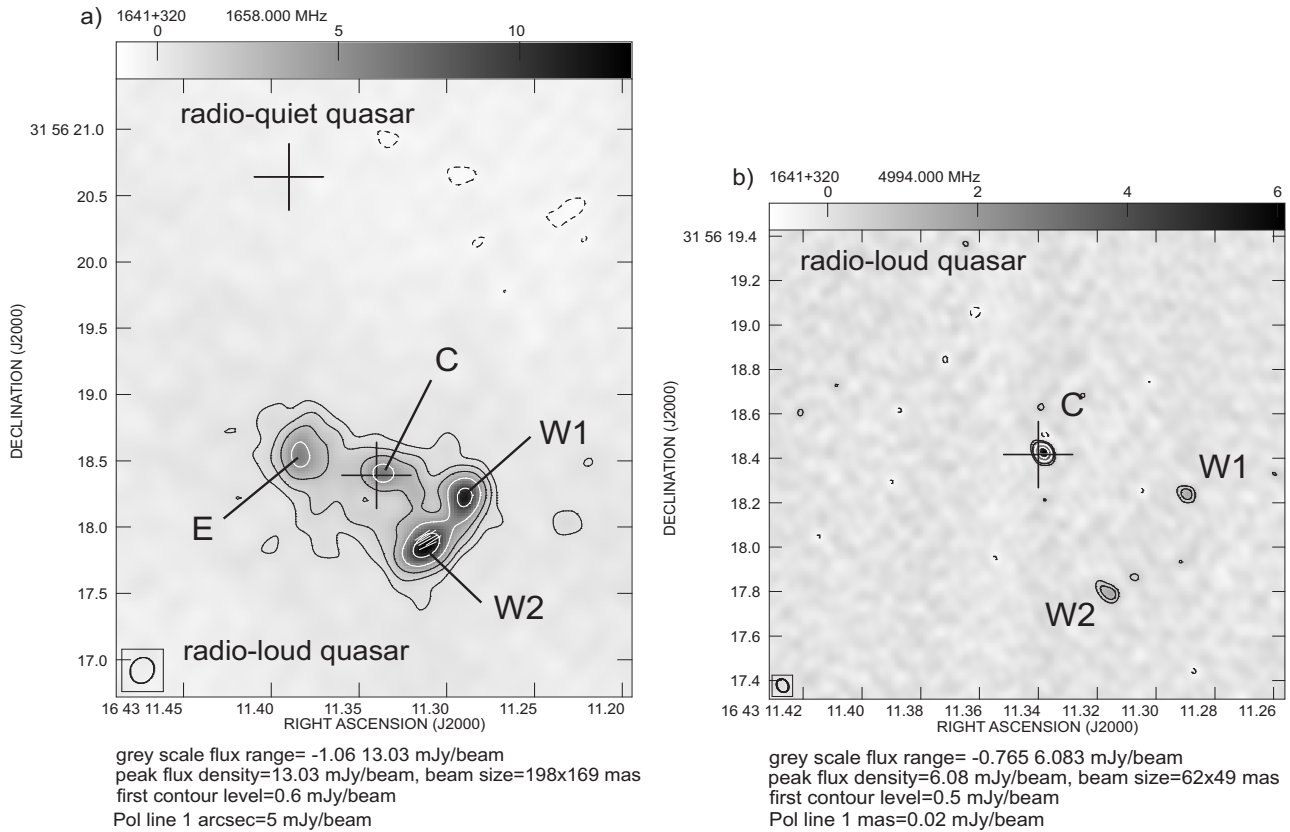}
\caption{a) The 1.66\,GHz MERLIN radio image, b) The 5\,GHz MERLIN radio
image. Contours increase
by a factor 2, and the first contour level corresponds to $\approx
3\sigma$, vectors represent the polarized flux density.
A cross indicates the position of an optical object found using
the most actual version of SDSS. The shape of the beam is given in the 
left-bottom corner of each image.}
\label{pairs}
\end{figure*}

\section{Results}
\label{radio}

The 1.66\,GHz MERLIN radio image of FIRST\,J1643+3156
shows four components of the source (Fig.~\ref{pairs}a). A component
indicated as
C, which position is well correlated with the position of the
optical counterpart is a radio core. We have measured flux densities of the
indicated components (Table~\ref{component}) and calculated spectral index
for
component C which is $\alpha_{1.66\mathrm{GHz}}^{5\mathrm{GHz}}=0.6$.
However, this should be treated as an approximation since it is based on the
snapshot
observations were some flux can be missing. For this reason we did not
calculate spectral indices for weaker components: the radio lobes 
E, W1 and W2. However, in this case the difference between the flux
densities
of the components at both frequencies is so large that we can assume they
have steep spectra (Table~\ref{component}). The lobe E is the
weakest one in the 1.66\,GHz image and there is no trace of it in the
5\,GHz image (Fig.~\ref{pairs}b). The 1.66\,GHz MERLIN image shows that there is a connection 
between the component C and the south-western component W2. We
suggests that this could be the most current jet direction and particle
ejection.
There is also an extended emission around the four compact components in the
1.66\,GHz image. However, no extended emission and potential jet is visible
in the 5\,GHz image, and all three features detected in 5\,GHz image are very weak. 
W2 is also the only one component   
showing polarization and only at the frequency of 1.66\,GHz. 
Less then 1\% of the total emission of W2 is polarized at 1.66\,GHz. The
polarization angle amounts to $-59^{\rm o}\pm 15^{\rm o}$ (from N to E). 
The polarization vectors are often
perpendicular to the jet direction, as in this case, suggesting that the
component W2 is powered by the jet. Although the uncertainty of
polarization angle is large and without more sensitive 5\,GHz observations
of FIRST\,J1643+3156, the interpretation should be treated as estimation.

Using the SDSS spectrum of FIRST\,J1643+3156 we have measured its emission lines
widths and luminosities \citep{kun10b}. FIRST\,J1643+3156 has the largest
[O\,III] luminosity (L$_{[\rm O\,III]} = 1.1 \times 10^{43}~{\rm erg~s^{-1}}$) 
among the Low Luminosity Compact (LLC) objects we
observed and was classified as High Excitation Galaxy (HEG). 
Among the LLC objects which we consider as the candidates to the short-lived
radio objects population, we found three binary systems: the two galaxy pairs
(0854+210, 1506+354), and quasar binary FIRST\,J1643+3156 (1641+320). Their morphologies are
highly disrupted \citep{kun10a} and do not resemble typical CSS sources with symmetric
double lobes.  

\subsection{Optical Imaging}

\begin{figure}
\centering
\includegraphics[width=8cm,height=9cm]{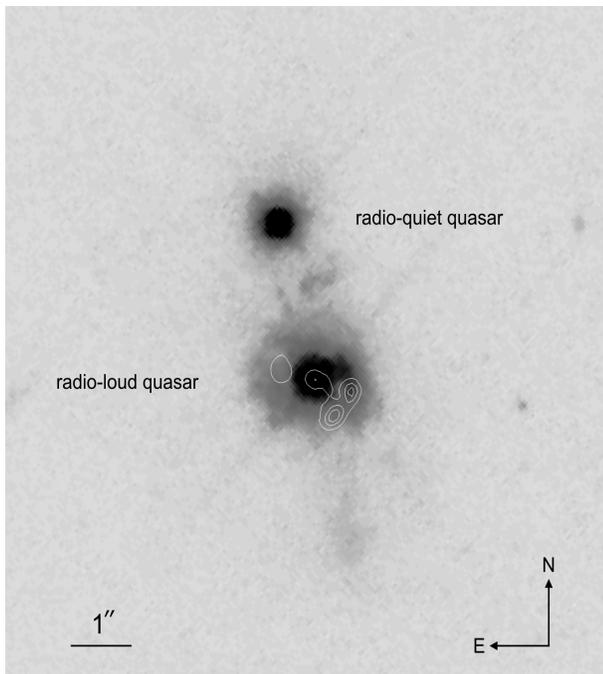}
\caption{The HST/ACS image of FIRST\,J1643+3156 in F814W filter with an
overlay of contours
of the radio emission at 1.66\,GHz. The two
bright quasar nuclei are visible. The host galaxy of the radio-loud quasar
is perturbed.}
\label{optical}
\end{figure}

FIRST\,J1643+3156 has been observed with the Wide Field Channel (WFC) of the Advanced Camera 
for Surveys (ACS) of the Hubble Space Telescope (HST) by \citet{ford} and
described by \citet{martel}. The images were made in 2005 in three filters:
{\it g}\,(F475W), {\it r}\,(F625W), and {\it I}\,(F814W), and the exposure
time of the observations was
$\sim$18 minutes in {\it g} and {\it r} band, and $\sim$36 minutes in
{\it I} band.
We have retrieved the data from the HST Archive and present the {\it I} band
image of FIRST\,J1643+3156 in Figure~\ref{optical}. As described by 
\citet{martel} the host galaxy of the radio-loud quasar is 
highly disturbed while the host galaxy of the radio-quiet source appears
smooth and unperturbed. Three morphological features can be distinguished in
the image of the host galaxy of the radio-loud quasar: a bright arc extending
$\sim$3\,kpc west of the nucleus, a large diffuse knot located $\sim$8\,kpc to the
south, and a group of
filaments to the north-east of the nucleus. Distorted filaments are also
observed between the radio-loud and radio-quiet quasars. 
We have made an overlay of the contours
of the radio emission at 1.66\,GHz on the optical image. We suggest that the
bright arc on the optical image corresponds to the disturbed radio structure
what may indicate they are both an effect of the quasars' interaction.

\section{Discussion}
\label{dis}
FIRST\,J1643+3156 belongs to the population of Compact Steep Spectrum (CSS) sources.
CSS sources form a well defined class of compact radio objects ($\leq$
20~kpc) and are considered to be younger progenitors of large radio-loud
AGNs.  This interpretation of
the CSS class has now become part of a standard model \citep{f95},
and \citet{r96} have proposed an evolutionary scheme
unifying the three classes of radio-loud AGNs: Gigahertz-Peaked Spectrum
(GPS) sources, CSS sources and large scale radio objects.  
Two pieces of evidence definitely point towards GPS/CSS sources being young
objects: lobe proper motions (up to 0.3c) giving kinematic ages as low as
$\sim$$10^3$~years for GPSs \citep{ocp98,gir03,pol03} and
radiative ages typically $\sim$$10^5$~years for CSSs \citep{mur03}.
Although these AGNs are small-scale objects, in some cases CSO/GPS sources
are associated with much larger radio structures that extend out to many
kiloparsecs. In these cases, it has been suggested that the CSO/GPS stage represents a
period of renewed activity in the life cycle of the AGN \citep{stan05,kun06}.
\citet{rb97} have also proposed a model in which extragalactic
radio sources are intermittent on timescales of $10^{4}-10^{5}$ years,
suggesting that many young AGNs, CSS and GPS sources, might be short-lived
objects. Detection of several candidates for dying compact sources
\citep{gir05,kun06, kun10a, orienti} supports this view.

The ignition of the AGN activity can be caused by the major merger or
instabilities in the accretion flow. In the second case it is highly
probable that the activity will be intermittent \citep[e.g.][]{czerny09},
although this can change when considering non-standard situation, namely the binary
system (see section 4.1).
Observations indicate, that about 50\% of young AGN contain double
nuclei in their host galaxies or exhibit morphological distortions
that are supposed to be due to the past merging events
\citep{odea98,liu04, tao, kunradio}. 

The CSS object FIRST\,J1643+3156 is an uncommon young AGN because of its complex
radio morphology and fact, that it is a part of a binary quasar.
The disturbed radio and optical structure of FIRST\,J1643+3156 suggests that 
both quasars in this system
are physically bound and influence each other during the evolution cycle
(Figure~\ref{optical}).
However, there is not enough information to determine the geometry of the
binary system: the full orbital velocity, radius and period, which is
important in explaining the distortions in radio structure we observed. 
Based on the observed line-of-sight velocity difference between the binary
system components ($V= 300$ km s$^{-1}$)  and its separation ($d=$ 15\,kpc), we
can roughly estimate the timescale of the tidal perturbation between the two
quasars as $T_{\rm p}= d/V = 5 \times 10^{7}$ years. The age of the radio source can be
estimated e.g. from the size {\it l} of the source, as $T_{\rm a}= l/v_{\rm
lobe}$. Taking the
moderate lobe velocity of $v_{\rm lobe}=0.1$c and even the largest linear size of
$l=$ 10.56\,kpc (which is a distance between peaks of components E and W1
assuming they belong to the same phase of activity), the estimated age of
the source is on the order of $3\times10^{5}$ years. Thus we consider a
few 
scenarios which may explain the radio properties of the observed binary quasar.

\subsection{Intrinsic activity restart and its modulation by the companion}
\label{intrinsic}

The first scenario is based on the conclusion that if the radio activity
timescale ($\sim 10^{5}$ years) is much shorter than the tidal perturbation timescale
($\sim 10^{7}$ years), the radio activity of the radio-loud component does not have
to be induced directly by the interaction with the companion galactic core.
It could be the intrinsic mechanism that made the quasar start its
activity. The radio structure ${\rm E-C-W1}$ could be then  
the sign of the first period of activity which was prematurely ceased by the
instability of the accretion disk \citep{janiuk02, czerny09}. The model for the internal
instability in the accretion disk is based on the radiation pressure
dominance. This physical effect leads to the periodic outbursts of the disk
luminosity, intermittent with the low luminosity periods. The feature W2 is
then interpreted as the subsequent activity phase which could be now changed due to the
interaction with the companion galaxy. We have performed such modelling (see
Section ~\ref{modulation} for the full description and plots) using the
accretion disk instability cycle. To account for the influence of the
companion galaxy, we change the outer boundary condition in the evolutionary
calculations. We assumed the accretion rate to be a periodic function of
time and this perturbation is given by the tidal interaction in the period
$\sim 10^{7}$ years. The black hole mass was $M_{\rm BH} = 3\times
10^{8} M_{\odot}$, as estimated for the
radio-loud component \citep{shen10}. As a result we obtained that the 
interaction with the companion quasar influences the modulation and overall
evolution cycle of the primary quasar. It can dramatically change the
activity pattern and lead to the prolonged, up to a few million years,
activity outbursts. 

Below, we present the results of the numerical modeling of the 
intermittent activity for the binary quasar. 

\subsubsection{Luminosity outbursts of the core of radio-loud component}
\label{outbursts}

\begin{figure}
\includegraphics[width=8cm,height=8cm]{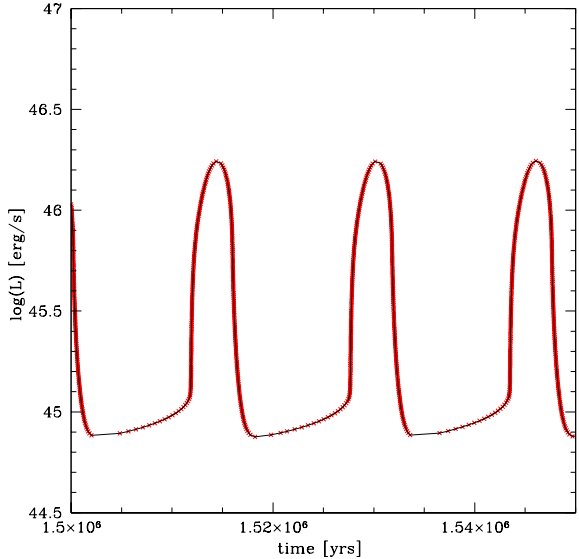}
\caption{Time evolution of the active core luminosity
in the cycle due to the radiation pressure instability.
The black hole mass is $M = 3\times 10^{8} M_{\odot}$,
external accretion rate is constant $\dot m_{\rm ext} = 0.14$ 
and viscosity parameter is $\alpha=0.01$}
\label{fig:cycle1}
\end{figure}

In Figure \ref{fig:cycle1} we show an exemplary lightcurve
that results from the accretion disk instability model. The
model
 was described in detail with appropriate time-dependent equations 
and numerical method in \citet{janiuk02}.

The luminous core in our example 
exhibits regular outbursts for the external (mean) accretion rate
of 14 per cent of the Eddington rate. The duration of the outbursts
and their 
separation depends on black hole mass and viscosity parameter. In this case,
the outbursts cycle lasts about $ 2\times 10^{4}$ years.

In this model, 
we postulate that a large fraction of the disk power at outburst is
transferred to the jet and provides a source for its kinetic energy,
parametrized by the fraction $\eta_{\rm jet}$ in the range (0;1).
This jet fraction is dependent on time and radius via the local
accretion rate, depending on both time and distance from the black hole:
\begin{equation}
\eta_{jet} = 1 - {1 \over 1+  A_{jet} \dot m(r,t)^{2}}.
\label{eq:ajet}
\end{equation}
with $A_{\rm jet}$ being a parameter of the model. The physical 
meaning of this parametrization is that at the Eddington accretion limit 
there would be equipartition between the jet and disk radiation for 
$A_{jet}=1$. We consider here the accretion rates well below the Eddington limit
and we adopt a moderate value of $A_{\rm jet}=2.5$.
The disk-jet coupling results in reducing the outbursts amplitudes in comparison to the models
without the jets \citep{jc11}. This model is plausible for the radio-loud quasar
and consistent with its observed luminosity.

\subsubsection{Influence of the companion radio-quiet quasar}
\label{modulation}

To account for the influence of the companion galaxy, we change the outer 
boundary condition
in the evolutionary calculations. The external accretion rate, instead of 
being a constant 
parameter, is
a periodic function of time:
\begin{equation}
\log {\dot m_{ext} \over \dot m_{0}} = \sin ( {2\pi \over T_{p}} t)
\label{eq:pert}
\end{equation}
where $T_{\rm p}$ is a characteristic perturbation timescale.
We assume that this perturbation is given by the tidal interaction and
$T_{\rm p} \approx 10^{7}$ years.
The accretion rate is parametrized by $\dot m_{0} = 0.014$
and changes from 0.1$\dot m_{0}$ to 10$\dot m_{0}$, so that it never 
exceeds the Eddington 
limit during the perturbation cycle.

\begin{figure}
\includegraphics[width=8cm,height=8cm]{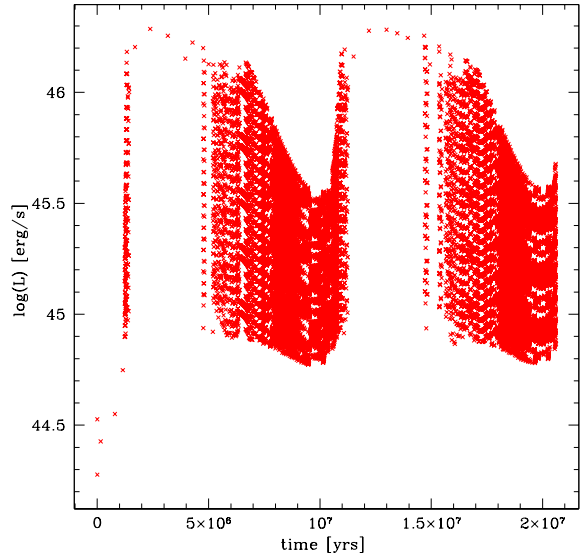}
\caption{The outburst amplitudes of the radio-loud component's core 
luminosity in the model with 
the accretion rate modulation due to the
binary interaction.
The perturbation is parametrized according to Eq.
\ref{eq:pert}, with $T_{p}=10^{7}$ years and $\dot m_{0} = 0.014$.
The viscosity is $\alpha=0.01$, black hole mass is
$M = 3\times 10^{8} M_{\odot}$  and
$A_{jet}=2.5$.} 
\label{fig:tper1}
\end{figure}

The modulation of the material supply rate to the outer disk edge results 
in periodic changes of the outburst pattern. Both the durations and
amplitudes are affected.

In Figure \ref{fig:tper1} we show the modulated evolution of the
unstable disk coupled with a jet. 
Note that the Figure \ref{fig:tper1} shows now much longer timescale
than in Figure \ref{fig:cycle1} and therefore the individual
shorter timescale outbursts are marked as a shaded regions.
The adopted black hole mass was 
$M=3\times 10^{8} M_{\odot}$, starting accretion rate 
$\dot m_{0} = 0.014$ and the jet parameter $A_{\rm jet}=2.5$.
Now, the pattern of the long term evolution is the following. The largest 
external accretion rates result in a persistent outburst state of the disk, 
while the luminosity
fluctuations occur at small accretion rates. This long persistent high state
was not obtained in the simulations with a constant boundary condition, and
is an effect of the modulation.
Moreover, it was not obtained in the models without the disk-jet coupling and
we had persistent quiescent states instead. 
Also, we noted that the transition rates
from the intermittent to the persistent state and vice versa, are not equal:
in the present simulation, 
the first transition occurs at $\dot m_{1}=0.08$ and the second transition at
$\dot m_{2}=0.02$.
This complex behavior results from the adopted jet cooling term, which
depends non-linearly on time and distance. In this way, the accretion disk 
plasma
holds the ``memory'' of the past conditions as the information propagates 
inwards
in the disk in the viscous timescale, so that
 we obtain an effect of 'hysteresis'.

In conclusion, the accretion rate modulation overimposed on the cyclic 
outbursts
influences the source behavior on long 
timescales. This modulation is due to the companion galaxy tidal forces, which
influence the rate of matter supply to the outer regions of the accretion disk.
The transitions from the strong, periodically active behavior,
through the 'flickering', and then to the persistent high state of the source may occur
due to the companion interaction. 

In the quasar FIRST\,J1643+3156 we may therefore be observing
the active core that accretes near the transition rate from the
intermittent to the persistent high state. 
We consider here a scenario in which the radio structure ${\rm E-C-W1}$ is a sign
of the past activity, and W2 is a current activity episode that can be prolonged
up to a few million years, due to the interaction with the companion quasar.

We emphasize here that the above numerical simulations give a novel result and 
to our knowledge such hydrodynamical instability models
with a time-dependent boundary condition have not been performed before.
The details of the simulations are depending also on 
the presence or absence of the jet or outflow from the disk 
and are intended to be discussed elsewhere (Janiuk et al. in preparation). 

The instability scenario discussed above explains the episodic activity of the radio source 
(sec. \ref{outbursts}) as well as the prolonged activity episodes due to the influence of the 
companion (sec. \ref{modulation}). In this way 
we accounted for the timescales appropriate for the observed source. However, 
because the geometry of the structures observed
 is also complex and the instability
scenario itself is not able to predict geometrical changes (the
non-colinearity of the
observed structures), below we discuss
further possibilities.

\subsection{Geometry changes due to the interaction with the companion}
\label{interaction}

The intermittent radio activity can be caused by an intrinsic mechanism,
as well as modulated or directly triggered by the interaction with 
the companion galaxy.

\subsubsection{Precession}
The observed complex radio structure of FIRST\,J1643+3156 could be then caused by the
accretion disk precession due to tidal torques induced by the companion in
the binary system \citep{caproni}. In this case,
the precession due to the tidal interaction in a binary system
will occur if the binary orbit is nor coplanar with the accretion disk.
Considering this scenario the disturbed radio morphology can be explained in
the following way. The structure ${\rm E-C-W1}$ is the first episode of the
radio source activity and the first direction of the radio jet and particle
ejection. Then the precession caused the change in the accretion disk plane
and the new jet direction of FIRST\,J1643+3156
is the south-western one - the jet is fuelling the component W2.
This precession scenario is therefore independent but also 
complementary to the
episodic activity scenario discussed above in Sec. 4.1, as it explains
the lack of colinearity between the subsequent structures
(to account simultaneously for both effects, i.e. the timescales and 
geometry changes, would require essentially 3D time-dependent global modeling of a 
non-axisymmetric unstable accretion disk, which is beyond the scope of the 
present work).
The estimated age of the radio structure of FIRST\,J1643+3156 is on the
order of $10^{5}$ years. The timescale of the
decay of the components no longer powered by the jet (lobes E and W1 ?) is
comparable with the
timescale of its activity phase. However, during the first few periods of
inactivity, the radio luminosity fades rapidly \citep{rb97} making the
potential distortions well visible only during a short period of time.
The fading source structure is then characterized by weak emission
detectable only in deep, low-frequency observations.
It is possible then that the radio observations of FIRST\,J1643+3156 were
made just in time to detect the distortions at lower 1.66\,GHz frequency.

\subsubsection{Jet-ISM interaction}
\label{jet}

Finally, we consider a scenario in which the radio activity of the
FIRST\,J1643+3156 has been
directly triggered by the interaction with the companion galaxy but its   
radio structure is a result of the jet interaction with the host galaxy
environment.
The 1.66\,GHz image of FIRST\,J1643+3156 (Fig.~\ref{pairs}a)
shows a diffuse radio emission around the radio-loud quasar
suggesting very dense medium of the host galaxy of the source and
indicating strong interactions. 
Simulations of
colliding disk galaxies \citep{barnes} tracked the evolution of both gas
and stars in the merger during the subsequent orbital loops. The gas
accumulates in the nucleus of each galaxy as the two orbit away from each
other, but stronger perturbation of the incoming galaxies are visible only
after the first pericentric passage. In such an
environment, a black hole may undergo many fuelling events and each event
may completely disrupt and/or restart the jet. After each renewal, the jet
may need to force its way through the changing nuclear environment anew.
In the case of FIRST\,J1643+3156, the features W1 and W2 could be parts of
the same jet. The jet is changing its orientation during propagation in the
central regions of the host galaxy due to interactions with the dense
environment. The feature E could then be the counter-jet. 
The presence of strong [O\,III]$\lambda 5007$ in FIRST\,J1643+3156
\citep{broth, kun10b} suggests
the jet-ISM interactions. 
According to \citet{labiano}, the expansion of the radio
source through the host ISM could be triggering or enhancing the
[OIII]$\lambda 5007$ 
line emission through direct interaction.
The presence of the cold accreting material could be responsible for even higher
gas excitation, and consequently for higher [O\,III]$\lambda 5007$ line
emission. It has been discussed by \citet{butti} that the High Excitation Galaxies
(HEGs), like the FIRST\,J1643+3156, are powered by accretion of cold gas
provided probably by the merger with a gas rich galaxy.

\section{Summary}

FIRST\,J1643+3156 is an unusual and statistically very rare, low redshift binary
quasar. 
The radio morphology of FIRST\,J1643+3156 is complex
on both frequencies and consists of four components which indicates
rapid change of jet direction and/or restart of the jet. 
The host galaxy of the radio-loud quasar is also highly disturbed 
and we suggest that the first pericentric passage took place in this pair 
igniting and/or changing the radio activity and morphology in the radio-loud
component. We discussed several possible scenarios that
could explain such a complex radio morphology:
(i) accretion disk instability, modulated by the interaction with the companion, 
(ii) precession of the disk/jet due to the companion's tidal forces 
and (iii) jet-cloud interactions.

New observations could test the evolution scenarios presented here. A
more sensitive radio observations may allow to trace the jet pattern and the
change of its direction at both 1.66\,GHz and 5\,GHz frequencies. 
An infrared and X-ray observations could give us information about the
intrinsic absorption and potential presence of the dense environment in
FIRST\,J1643+3156.

Based on the analysis described above we suggest that 
regardless of the mechanism that caused 
the distortions of the radio structure visible in the
binary system, they may be short-lived phenomena. The disturbed radio components that are no
longer powered by the jet can quickly fade away. This makes the potential distortions 
detectable only during a short period of time and implies a low detection rate. 
To speculate what could be the past and future history of FIRST\,J1643+3156
binary system, we performed numerical simulations concerning the 
influence of the companion quasar to the activity of the radio-loud one. 
The results show that the ignited or subsequent
activity phase can be prolonged up to a few million years and limits the
occurrence of distorted radio structures, at least those caused by the internal
instability in the accretion disk.

\acknowledgments
We thank B.Czerny, M. Sikora and M. Gawro\'nski for helpful discussions.\\
We thank an anonymous referee for careful reading of the manuscript and 
constructive comments.\\
MERLIN is a UK National Facility operated by
the University of Manchester on behalf of STFC.\\
We used the observations made with the NASA/ESA Hubble Space Telescope,
obtained from the data archive at the Space Telescope Science Institute.
STScI is operated by the Association of Universities for Research in
Astronomy, Inc. under NASA contract NAS 5-26555.\\
This research has made use of SAOImage DS9, developed by Smithsonian
Astrophysical Observatory.\\
Part of this work was supported by the:
{\it COST Action MP0905 Black Holes in a Violent Universe}.\\
M.K-B. acknowledges support from the Polish Ministry of Science and Higher
Education under grant N N203 303635.

\end{document}